\documentclass{ws-procs975x65}
\usepackage{graphicx}

\def\beq{\begin{equation}}
\def\eeq{\end{equation}}
\begin{document}

\title{Sterile neutrino dark matter and core-collapse supernovae}
\author{Grant J. Mathews$^*$ and MacKenzie Warren}

\address{Department of Physics, University of Notre Dame,\\
Notre Dame, IN 46556, USA\\
$^*$E-mail: gmathews@nd.edu}

\author{Jun Hidaka and Toshitaka Kajino} 

\address{National Astronomical Observatory of Japan\\
Mitaka, Tokyo 181-8599, Japan}

\begin{abstract}
We have explored the impact of sterile neutrino dark matter on core-collapse supernova explosions. We have included oscillations between electron neutrinos or mixed $\mu,\tau$ neutrinos and right-handed sterile neutrinos into a supernova model. We have chosen sterile neutrino masses and mixing angles that are consistent with sterile neutrino dark matter candidates as indicated by recent x-ray flux measurements. Using these simulations, we have explored the impact of sterile neutrinos on the core bounce and shock reheating. We find that, for ranges of sterile neutrino mass and mixing angle consistent with most dark matter constraints, the shock energy can be significantly enhanced and even a model that does not explode can be made to explode. In addition, we have found that the presence of a sterile neutrino may lead to detectable changes in the observed neutrino luminosities.
\end{abstract}

\keywords{supernova neutrinos, dark matter theory, core-collapse supernovae}

\bodymatter

%%%%%%%%%%%%%%%%% now a standard article style for the most part

\section{Introduction}

Sterile neutrinos, a proposed fourth neutrino flavor, are one viable dark matter candidate.\cite{abazajian2001b}  A sterile neutrino is an electroweak singlet and is thus consistent with limits from the {\it LEP} measurement \cite{LEP2006} of the width of the $Z^{0}$ gauge boson.  The standard model does not provide any predictions about this proposed particle, but  bounds can be placed using astronomy, cosmology  and supernovae\cite{abazajian2001a,abazajian2001b,boyarsky2006a,boyarsky2006b,boyarsky2007,boyarsky2009b,boyarsky2009c,boyarsky2014,bulbul2014,chan2014} on the mass and mixing angle parameter space.  If sterile neutrino dark matter can radiatively decay,\cite{pal1982}  x-ray observations\cite{boyarsky2006b,boyarsky2007,boyarsky2006a,boyarsky2009b,boyarsky2009c,boyarsky2014,bulbul2014} from galaxies and galaxy clusters can also be used to place bounds, i.e $1\text{ keV} < m_{s}< 18 \text{ keV}$ and $\sin^{2} 2 \theta_{s} \lesssim1.93 \times 10^{-5} \left({m_{s}}/{\text{keV}}\right)^{-5.04}$.

The explosion mechanism of CCSNe remains an outstanding problem in physics as well.  Spherically symmetric models do not easily explode and  two-dimensional  and three-dimensional models that do explode often have too little energy to match observations.\cite{janka2012}  The problem remains that, although a shock forms successfully and propagates outward in mass, it loses energy to the photodissociation of heavy nuclei and becomes a standing accretion shock.  

However, even one dimensional models explode in simulations with enhanced neutrino fluxes, either from convection below the neutrinosphere \cite{wilson1988,wilson1993,book} or a QCD phase transition.\cite{fischer2011}  Along this vein, we have explored the resonant mixing between sterile and electron neutrinos (or antineutrinos) to increase the early neutrino luminosity and revitalize the shock.\cite{hidaka2006,hidaka2007,warren2014,warren2016}

Hidaka and Fuller \cite{hidaka2006,hidaka2007} were the first to propose that sterile neutrinos could serve as an efficient neutrino energy transport mechanism in the supernova core.  They used a one zone collapse calculation to study the resonant oscillations of a sterile neutrino with the mass and mixing angle of a warm dark matter candidate. They found that the resonant mixing of electron and sterile neutrinos can serve as an efficient method of transporting neutrino energy from the protoneutron star core, where high energy neutrinos are trapped, to the stalled shock to assist in neutrino reheating.  This mechanism is highly sensitive to the feedback between neutrino oscillations and the local composition, energy transport, and hydrodynamics and warranted detailed numerical studies.\cite{warren2014,warren2016}

In Refs.~\refcite{warren2014} \& \refcite{warren2016} coherent active-sterile neutrino oscillations were studied using the University of Notre Dame-Lawrence Livermore National Laboratory (UND/LLNL) code,\cite{book,bowers1982} a spherically symmetric general relativistic supernova model with detailed neutrino transport and hydrodynamics. The impact on shock reheating of sterile neutrinos with masses and mixing angles consistent with dark matter constraints was studied.  Sterile neutrino dark matter candidates can enhance the shock energy and lead to a  successful explosion, even in a simulation that would not otherwise explode.\cite{warren2014}

\section{Matter-enhanced Neutrino Oscillations\label{sec:osc}}

With the inclusion of a sterile neutrino, the full neutrino mixing problem requires a complete understanding of all mass differences and mixing angles in the $4\times 4$ mixing matrix.  However, for this work, only two neutrino mixing between electron neutrinos $\nu_{e}$ and sterile neutrino $\nu_{s}$ (or their antiparticles) have been considered.  This is sufficient for exploring the impact on the explosion energy since electron neutrinos and antineutrinos dominate in the gain region during shock reheating.

Matter-enhanced neutrino oscillations in supernovae occur via the Mikheyev-Smirnov-Wolfenstein (MSW) effect.\cite{mikheyev1985,wolfenstein1978}  As neutrinos propagate through matter, they experience an effective potential from charged and neutral current interactions due to forward scattering on baryonic matter, electrons, and other neutrinos. Each neutrino flavor will experience a different potential because $\nu_{e}$ experience both charged and neutral current interactions whereas $\nu_{s}$ do not experience any weak interactions. This can induce a coherent effect where maximum mixing is possible, even for a small vacuum mixing angle, when the phase arising from the potential difference cancels the phase due to the mass difference.  The forward scattering potential experienced by electron neutrinos is
\begin{equation}
V(r) = \frac{3 \sqrt{2}}{2} G_{F} n_{B} \left(Y_{e} + \frac{4}{3} Y_{\nu_{e}} + \frac{2}{3} Y_{\nu_{\tau}} -1\right)~,
\end{equation}
where $G_{F}$ is the fermi coupling constant, $n_{B}$ is the baryon number density, and $Y_{i}$ is the number fraction of species $i$.  The antineutrino species will experience forward scattering potentials with the opposite sign.  In the supernova environment, one can assume $Y_{\nu_{\mu}} = Y_{\nu_{\tau}} = 0$, since $\nu_{\mu}$ and $\nu_{\tau}$ neutrinos and antineutrinos are produced via pair emission processes and thus occur in equal numbers.

This results in an ``effective'' in-medium mixing angle,\cite{warren2014,warren2016}
\begin{equation}
\sin^{2} 2 \theta_{M} (r) = \frac{\Delta^{2} \sin^{2} 2 \theta_{s}}{(\Delta \cos 2 \theta_{s} - V(r))^{2} + \Delta^{2} \sin^{2} 2 \theta_{s}}~.
\end{equation}
From this expression, it is simple to see that one can achieve maximal mixing in matter, even for small vacuum mixing angles.  Such resonant mixing will occur for neutrinos with the energy where $\sin^{2} 2 \theta_{M} = 1$,
\begin{equation}
E_{res} = \frac{\Delta m^{2}}{2 V(r)} \cos 2 \theta_{s}~.
\end{equation}

In this work, only coherent and adiabatic oscillations are considered.\cite{warren2014,warren2016}  This ensures that all electron neutrinos of the given flavor with the resonance energy $E_{res}$ will oscillate to sterile neutrinos, and vice versa.\cite{mikheyev1985,wolfenstein1978}  It is probable that incoherent, scattering-induced oscillations will be significant in the high matter densities of the central core, but this will not be a dominant effect and will be left to future work.

\section{Results\label{sec:results}}

A model that can successfully explode was used as a baseline for comparison.  The UND/LLNL supernova model\cite{bowers1982,book} is a spherically symmetric, general relativistic hydrodynamic supernova model. We have used the $20 \text{ M}_{\odot}$ progenitor from Woosley \& Weaver \cite{woosley1995}.

\begin{figure}[h]
   \centering
   \includegraphics[width=3.0in]{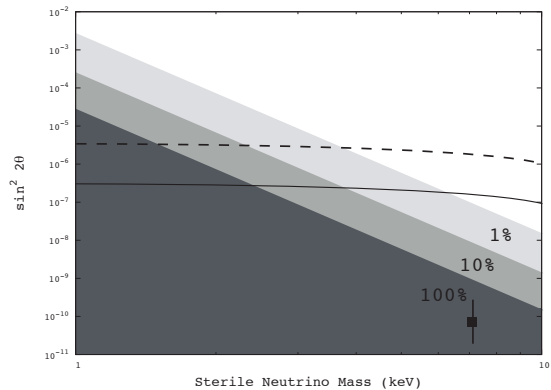} % requires the graphicx package
   \caption{Sterile neutrino mass $m_{s}$ and mixing angle $\sin^{2} 2 \theta_{s}$ parameter space.  The region above the solid line enhances the supernova explosion energy by $1.01\times$ compared to a simulation without a sterile neutrino. The region above the dashed line enhances the explosion energy by $1.1\times$.  The dark gray shaded region shows the parameter space allowed by x-ray flux measurements if sterile neutrinos make 100\% of the observed dark matter mass.\cite{abazajian2001a,boyarsky2006a,boyarsky2006b,boyarsky2007,boyarsky2009c}  The medium gray region is for 10\% of the observed dark matter mass and the light gray region is for 1\% of the observed dark matter mass.  The solid black square shows the most recent best fit point from the x-ray flux from Boyarsky et al\cite{boyarsky2014} and Bulbul et al.\cite{bulbul2014}  Figure from Warren et al.\cite{warren2016}}
   \label{fig:parameter}
\end{figure}

Figure~\ref{fig:parameter} shows the enhancement to the explosion energy in a simulation with a sterile neutrino compared to a simulation without a sterile neutrino.  Sterile neutrino masses were considered in the range $1 \text{ keV} <  m_{s} < 10 \text{ keV}$ and mixing angles in the range of $10^{-11} < \sin^{2} 2\theta_{s} < 10^{-2}$, which includes the region that corresponds with dark matter candidates.  The shaded regions show the parameter space allowed for sterile dark matter by recent x-ray flux measurements.\cite{abazajian2001a,boyarsky2006a,boyarsky2006b,boyarsky2007,boyarsky2009c}  There is a large region of the parameter space that both enhances the explosion energy of core-collapse supernovae and is satisfies constraints on sterile neutrino dark matter.

To illustrate how the explosion energy enhancement occurs, consider a  single choice of sterile neutrino mass $m_{s} = 1$ keV and mixing angle $\sin^{2} 2\theta_{s} = 10^{-5}$.  Any mass and mixing angle that causes an enhancement will show similar behavior.

The enhancement to the kinetic energy isn't evident until $\sim0.2$s post-bounce, when neutrino reheating becomes important.  By 1s post-bounce, the explosion energy is enhanced by a factor of $2.2\times$ compared to the simulation without a sterile neutrino.

This dramatic enhancement to the kinetic energy is due to increased neutrino heating in the simulation with a sterile neutrino.  The presence of a sterile neutrino enhances the luminosities of all neutrino and antineutrino species considered here.  The  enhancement to the neutrino luminosities does not become significant until $\sim 0.1$s post-bounce, which is when oscillations to a sterile neutrino become important\cite{warren2014}.  The luminosities of all neutrino and antineutrino species are enhanced by  about $2\times$ until 0.4s post-bounce, which corresponds to the enhancement in the kinetic energy. The increased neutrino luminosities emerging from the protoneutron star increase the rate of neutrino heating in the gain region and lead to the enhanced explosion energy.

Although oscillations are only allowed between electron neutrinos $\nu_{e}$ and sterile neutrinos $\nu_{s}$ (and their antiparticles), enhancements are seen in the luminosities of all neutrino species.  This is because the oscillations $\nu_{e} \leftrightarrow \nu_{s}$ leads to a ``double reheating'' scenario:  the $\nu_{e} \leftrightarrow \nu_{s}$ oscillations cause additional heating near the location of the neutrinosphere, which causes increased neutrino cooling in all flavors, and finally these enhanced neutrino luminosities reheat the stalled shock.

%\begin{figure}[h]
%\centering
%\includegraphics[width = 0.5\textwidth]{lum_nue.pdf}
%\includegraphics[width = 0.5\textwidth]{lum_anue.pdf}
%\includegraphics[width = 0.5\textwidth]{lum_mu.pdf}
%\caption{Neutrino luminosity versus time post-bounce.  The dashed line is for a simulation without a sterile neutrino and the solid line is for a sterile neutrino with mass $m_{s} = 1$~keV and mixing angle $\sin^{2} 2 \theta_{s} = 10^{-5}$.  (a) shows the electron neutrino $\nu_{e}$ luminosity, (b) the electron antineutrino $\bar{\nu}_{e}$ luminosity, and (c) the $\mu-\tau$ neutrino $\nu_{\mu\tau}$ luminosity.  Figures from Warren et al.\cite{warren2016}}
%\label{fig:lum-e}
%\end{figure}

 In the simulation with a sterile neutrino, the neutrinosphere radius increases by about $1.4\times$ between 0.1s and 0.4s post-bounce, which corresponds with the increased neutrino luminosities.  The neutrinosphere radius is increased because the oscillations heat the protoneutron star surface by depositing energy at the location of the $\nu_{s} \rightarrow \nu_{e}$ resonance.  However, this additional heating of the protoneutron star surface does not increase the temperature of the neutrinosphere, but instead causes it to expand outward.  
 %Figure~\ref{fig:nutemp} shows the temperature versus radius at 250ms post-bounce for simulations with and without a sterile neutrino.  
 Although the location of the neutrinosphere is increased by $\sim1.4\times$, the temperature at the neutrinosphere is roughly the same in both simulations.  Thus the larger neutrinosphere radius leads to enhanced emission of neutrinos, but the neutrinos in both simulations have roughly the same characteristic temperature.

%\begin{figure}[h]
%\centering
%\includegraphics[width = 3.0 in]{nutemp.pdf}
%\caption{Temperature versus radius at 250ms post-bounce. The dashed line is for a simulation without a sterile neutrino and the solid line is for a sterile neutrino with mass $m_{s} = 1$~keV and mixing angle $\sin^{2} 2 \theta_{s} = 10^{-5}$.  The vertical lines mark the locations of the neutrinospheres for the respective simulations.  Figure from Warren et al.\cite{warren2016} }
%\label{fig:nutemp}
%\end{figure}

\section{Conclusions\label{sec:conc}}

Recent observations of galaxies and galaxy clusters indicate an unidentified emission line at $\sim3.5$keV.  This line may be due to the radiative decay of sterile neutrino dark matter with  $m_{s} \approx 7$keV.  If this is the case, bounds can be placed on the sterile neutrino mass and mixing angle from the observed photon energies and fluxes.  Further observations are needed to confirm the presence of this line in additional dark matter dominated environments, such as dwarf spheroidal galaxies.

For oscillations between an electron neutrino and sterile neutrino, a large region of  the sterile neutrino mass and mixing angle parameter space that is allowed by these observations leads to an enhancement of the explosion energy in core-collapse supernovae.  The enhancement is due to increased neutrino heating in the gain region caused by increased neutrino luminosities of all neutrino and antineutrino flavors.  The neutrino luminosities are enhanced due to a ``double reheating'' mechanism in the protoneutron star, where the surface of the protoneutron star is heated due to the oscillations between electron and sterile neutrinos, and in turn, the heating of the protoneutron star increases the luminosities of all neutrino and antineutrino flavors.

\section*{Acknowledgments}

Work at the University of Notre Dame supported by the U.S. Department of Energy under Nuclear Theory Grant DE-FG02-95-ER40934. One of the authors (MW) is also supported by U.S. National Science Foundation through the Joint Institute for Nuclear Astrophysics (JINA) Frontier center. This work was also supported in part by Grants-in-Aid for Scientific Research of the JSPS (20105004, 24340060).

\end{document}